\documentclass{ws-procs961x669}            

\begin{document}



 \cleardoublepage








\def\approxgt{\mathrel{\hbox{\rlap{\lower.55ex \hbox {$\sim$}}
        \kern-.3em \raise.4ex \hbox{$>$}}}}
\def\approxlt{\mathrel{\hbox{\rlap{\lower.55ex \hbox {$\sim$}}
        \kern-.3em \raise.4ex \hbox{$<$}}}}
\def \xmm {\emph{XMM-Newton} }
\def \sax {Beppo-SAX }
\def\pdot {\dot P}
\def\flux {\mbox{erg cm$^{-2}$ s$^{-1}$}}
\def\lum {\mbox{erg s$^{-1}$}}
\def\nh {$N_{\rm H}$}
\def\Omdot {\dot \Omega}
\def\ltsima{$\; \buildrel < \over \sim \;$}
\def\lsim{\lower.5ex\hbox{\ltsima}}
\def\gtsima{$\; \buildrel > \over \sim \;$}
\def\gsim{\lower.5ex\hbox{\gtsima}}
\def\msole{~M_{\odot}}
\def\msun{~M_{\odot}}
\def\mdot {\dot M}
\def\Msole{~M_{\odot}}
\def\Rsole{~R_{\odot}}
\def\rsun{~R_{\odot}}
\def\lsun{~L_{\odot}}
\def\Mdot {\dot M}
\def\hd {HD\,49798}
\def\rx {RX\,J0648.0--4418}
\def\hr {HD\,49798/RX\,J0648.0--4418}

\newcommand{\apx}[1]{^{\rm #1}}
\newcommand{\pdx}[1]{_{\rm #1}}

\newcommand{\apj}{{\it ApJ}}
\newcommand{\aj}{{\it AJ}}
\newcommand{\apjs}{{\it ApJS}}
\newcommand{\apjl}{{\it ApJ}}
\newcommand{\aap}{{\it A\&A}}
\newcommand{\aaps}{{\it A\&AS}}
\newcommand{\aapr}{{\it A\&A~Rev}}
\newcommand{\mnras}{{\it MNRAS}}
\newcommand{\araa}{{\it ARA\&A}}
\newcommand{\nat}{{\it Nature}}
\newcommand{\pasj}{{\it PASJ}}
\newcommand{\pasp}{{\it PASP}}
\newcommand{\apss}{{\it Ap\&SS}}
\newcommand{\physrep}{{\it Physics Reports}}


\title{The massive fast spinning white dwarf in the \hr\ binary }

\author{Sandro Mereghetti}
 
\address{IASF-Milano, INAF,\\
via A.Corti 12, Milano, 20133, Italy\\
 \email{sandro.mereghetti@inaf.it}}

\begin{abstract}
I  review the properties and discuss some of the puzzling aspects of the unique binary system composed of the luminous hot subdwarf HD 49798 and a white dwarf with mass of 1.2 $\msun$   and spin period of 13.2 s. 
This is one of the few massive white dwarfs with a dynamically measured mass and the one with the shortest spin period. 
It emits pulsed X-rays with a very soft spectrum, powered by accretion from the tenuous stellar wind of its companion of sdO spectral type. 
The current level of mass accretion cannot provide enough angular momentum to explain the small, but precisely measured, spin-up rate of 72 nanoseconds per year, which is instead best interpreted as the result of the radial contraction of this young white dwarf. 
The higher mass transfer rate expected during the future evolutionary stages of \hd\ will drive the white dwarf above the Chandrasekhar
limit, but the final fate, a type Ia SN explosion or the collapse to a millisecond pulsar, is uncertain.
\end{abstract}

\bodymatter

\section{Introduction}
\label{intro} 
 
The bright blue star \hd\ first attracted the interest of astronomers in the  sixties owing to the peculiarities of its spectrum  with prominent  helium and nitrogen lines and showing radial velocity variations, which indicated the  presence of a companion star\cite{jas63}. It was also noticed that its high galactic latitude (b=19$^{\circ}$)   implied a  distance above the Galactic disk unusually large for a main sequence OB star.

A few years later,   radial velocity measurements\cite{tha70} confirmed the binary nature of \hd\  with   the discovery of its orbital period (1.55 days) and the determination of the     mass function (0.27 $\msun$). 
However, no spectral or photometric signs of a second star   could be found, implying a faint low mass companion, such as a red dwarf or, possibly, a white dwarf. 
In 1994, the optical mass function was measured with higher precision\cite{sti94}
(0.263$\pm$0.004 $\msun$), but all the attempts to reveal the invisible companion of this single-lined
spectroscopic binary were unsuccessful.
  
The mystery of the companion star was solved, at least partially, when  soft X-ray emission with  a highly significant periodicity at 13.2 s was discovered\cite{isr97} from the $ROSAT$ source \rx , positionally coincident with \hd . This rapid periodicity can only be explained by the rotation of either a neutron star (NS) or a white dwarf (WD). In both cases, the expected optical emission  from these compact stars is much fainter than that of the much larger star \hd , and thus very difficult (or impossible) to be seen in optical photometric or spectral data.
 
Renewed interest in the \hr\ binary was prompted by the measurement of the X-ray mass function, obtained through  the timing analysis of the X-ray pulsations. 
This led to the determination of the masses of the two binary components, 
 $M_{HD}=1.50 \msun$ and
 $M_{X} =1.28 \msun$ for the subdwarf and  its companion, respectively\cite{mer09}.
As discussed in Sec.~\ref{WDNS}, there are now several arguments indicating that the companion of \hd\ is a massive WD,  with interesting implications for the future evolution of this system (Sec.~\ref{evol}).
   
\section{Properties of \hd\ }
\label{prop}

\hd\  (also known as CD-44 2920) is a bright blue star (B$-$V=$-$0.27) with apparent magnitude m$_V$=8.27, located in the southern sky 
(R.A.= 6$^{h}$ 48$^{m}$ 4.7$^{s}$, Dec.=$-$44$^{\circ}$ 18$'$ 58.4$''$, J2000). 
Early estimates indicated a distance of 650 pc, while  the   recent measurements obtained with Gaia\cite{gaiaDR3}  give   $d$=521$\pm$14 pc. 
Despite the slight downward revision of its  distance, \hd\ is still the most luminous  known subdwarf of O spectral type (sdO), 
with an absolute magnitude of $M_V$=$-$0.24 and a bolometric luminosity of 4.4$\times10^3$ $\lsun$.

A first application of non-LTE atmospheric models to the  spectrum of \hd\  yielded an  effective temperature of 
$T_{eff}$ = 47500$\pm$2000 K and a  surface gravity log\emph{g} = 4.25$\pm$0.2. The overabundance of helium,
that equals hydrogen in number (X$_{H}$=0.19 and X$_{He}$=0.78), was also confirmed\cite{kud78}.

More recently, accurate values for the element abundances and for the  stellar parameters of \hd\  were obtained with the analysis of high-resolution optical and UV spectra\cite{krt19} (see Table~\ref{tab-par} for some of these parameters). Note that the  \hd\ mass  derived from this atmosphere modelling is consistent with the dynamically determined value. 

A    particularly interesting property of \hd , in view of the X-ray emission from this binary, is the presence of a radiatively driven stellar wind, as it is clearly indicated by the P-Cygni profiles of nitrogen lines\cite{ham81}. The most recent estimates\cite{krt19}, that take into account   the peculiar abundances of \hd ,  give a mass loss rate of  2.1$\times$10$^{-9}$ $\msun$ yr$^{-1}$ and a wind terminal velocity $v_w$=1570 km s$^{-1}$. This wind  provides the matter that produces the accretion-powered X-rays from the compact  star. In addition, the stellar wind  is also responsible for the X-ray emission from  \hd\ itself\cite{mer13}, which is clearly visible, with a luminosity  $L_X\sim2\times10^{30}$ erg s$^{-1}$,   only when the hot subdwarf eclipses the much brighter X-ray flux emitted by its compact companion.  The relevance of \hd\ in the context of the study of the weak winds of early type subdwarfs\cite{krt16} is discussed in Ref.~\citenum{mer16r}.

 \begin{table}
\tbl{Main properties of the   binary system \hr\ }
{\begin{tabular}{@{}lcc@{}}
\toprule
 Parameter    &  Value & Reference  \\
\colrule

         $d$  [pc]           & 521$\pm$14 &           \citenum{gaiaDR3}     \\
      $P\pdx{orb}$   [d] &1.547666$\pm$0.000006&  \citenum{rig23} \\
           $P$  [s]           & 13.184246634$\pm7\times10^{-9}$ &  \citenum{rig23}  \\
        $\dot{P}$  [s s$^{-1}$]            &    (--2.28$\pm$0.02)$\times10^{-15}$ & \citenum{rig23}\\
        $i$ [deg]    & 84.5$\pm$0.7 & \citenum{rig23} \\
       $M_{WD}$ [$\msun$]     & 1.220$\pm$0.008    &   \citenum{rig23} \\
        $M_{HD}$  [$\msun$]    & 1.41$\pm$0.02             & \citenum{rig23}\\
$R_{HD}$  [$\rsun$]            & 1.08$\pm$0.06 &           \citenum{krt19,rig23}     \\
$T_{eff}$   [K]                  &  45900$\pm$800          &  \citenum{krt19} \\
log ($g$/1 cm s$^{-2}$)      & 4.56$\pm$0.08          &  \citenum{krt19} \\
$\epsilon_{He}$          & 0.74$\pm$0.07 &  \citenum{krt19} \\
log $\epsilon_{C}$          &  $<$--4.2 &  \citenum{krt19} \\
log $\epsilon_{N}$          & --3.1$\pm$0.2 &  \citenum{krt19} \\
log $\epsilon_{O}$          & --4.6$\pm$0.2 &  \citenum{krt19} \\
 $\dot M\pdx{w}$   [$\msun$  yr$^{-1}$]   &   $2.1\times10^{-9}$        &  \citenum{krt19} \\
$v_w$ [km s$^{-1}$]      &      1570     &  \citenum{krt19} \\
\botrule
\end{tabular}}
\begin{tabnote}
Abundances are expressed as  number density ratios relative to hydrogen ($\epsilon_X=N_X/N_H$)  \\
\end{tabnote}
\label{tab-par}
\end{table}

\section{X-ray properties}
\label{xray}

X-ray emission from the direction of \hd\ was   seen for the first time with the  {\it Einstein Observatory}  in March 1979, but the small number of counts in the HRI instrument (0.05-4 keV)  prevented a detailed analysis. The source was later observed with the {\it EXOSAT} satellite, but the detection in the CMA instrument was severely contaminated by the strong ultraviolet emission from \hd . 

A substantial  advance came with a {\it ROSAT} observation carried out in 1992, when the data of the PSPC instrument  revealed a very soft X-ray source with  flux of $\sim8\times10^{-13}$ erg cm$^{-2}$ s$^{-1}$  (0.1-2 keV), periodically modulated at 13.17 s \cite{isr97}. 
The fast pulsations, clearly resulting from the rotation of a compact star,   unequivocally indicate that the system contains a NS  or a WD. The poorly constrained spectral parameters derived with the PSPC were consistent with a large luminosity range,  from   $\sim10^{32}$ to more than 10$^{36}$   erg s$^{-1}$. Although it was not possible to distinguish between the NS and WD case based on the X-ray luminosity,   the presence of a coherent periodicity, which makes the system equivalent to a double spectroscopic binary,  opened  the possibility to obtain the masses of the two components.  

This was done thanks to a long  \xmm\ pointing, carried out in May 2008,  that was accurately scheduled to include also the orbital  phase of the expected  X-ray eclipse (never covered in previous X-ray observations). 
The  X-ray pulse delays induced by the orbital motion (giving the projected semi-major axis of the pulsar $a_X~{\rm sin} i$ = 4.2 $\rsun$) and the duration of the eclipse   (giving the system inclination 79$^{\circ}<i<84^{\circ}$) were measured\cite{mer09}. 
The derived value of $M_{X}$=1.28$\pm$0.05 $\msun$ was  still consistent with  both possibilities. However,  based on the X-ray luminosity of  $2\times10^{31}$   erg s$^{-1}$,  now well constrained by the good quality spectrum, the interpretation in terms of a massive WD was preferred\footnote{The same accretion rate onto a NS would give a luminosity higher by a factor $R_{WD}/R_{NS}\sim300$.}.
Updated system parameters,   derived from a phase-connected timing  analysis of  $ROSAT$,  \xmm\, and $NICER$ observations spanning  almost thirty years\cite{rig23}  are summarised in Table~\ref{tab-par}. 
They are slightly different from those obtained in previous works\cite{mer09}, mainly as a consequence of the revised distance.
In fact, the updated value of the radius of \hd\  leads to a higher inclination\footnote{The system inclination is derived from the observed  duration of the X-ray  eclipse (4311$\pm$52 s) and the known radius of the subdwarf.},
but main implications related to the masses of the two stars do not change significantly.

\begin{figure}[b]
\begin{center}
\includegraphics[angle=-90,width=11cm]{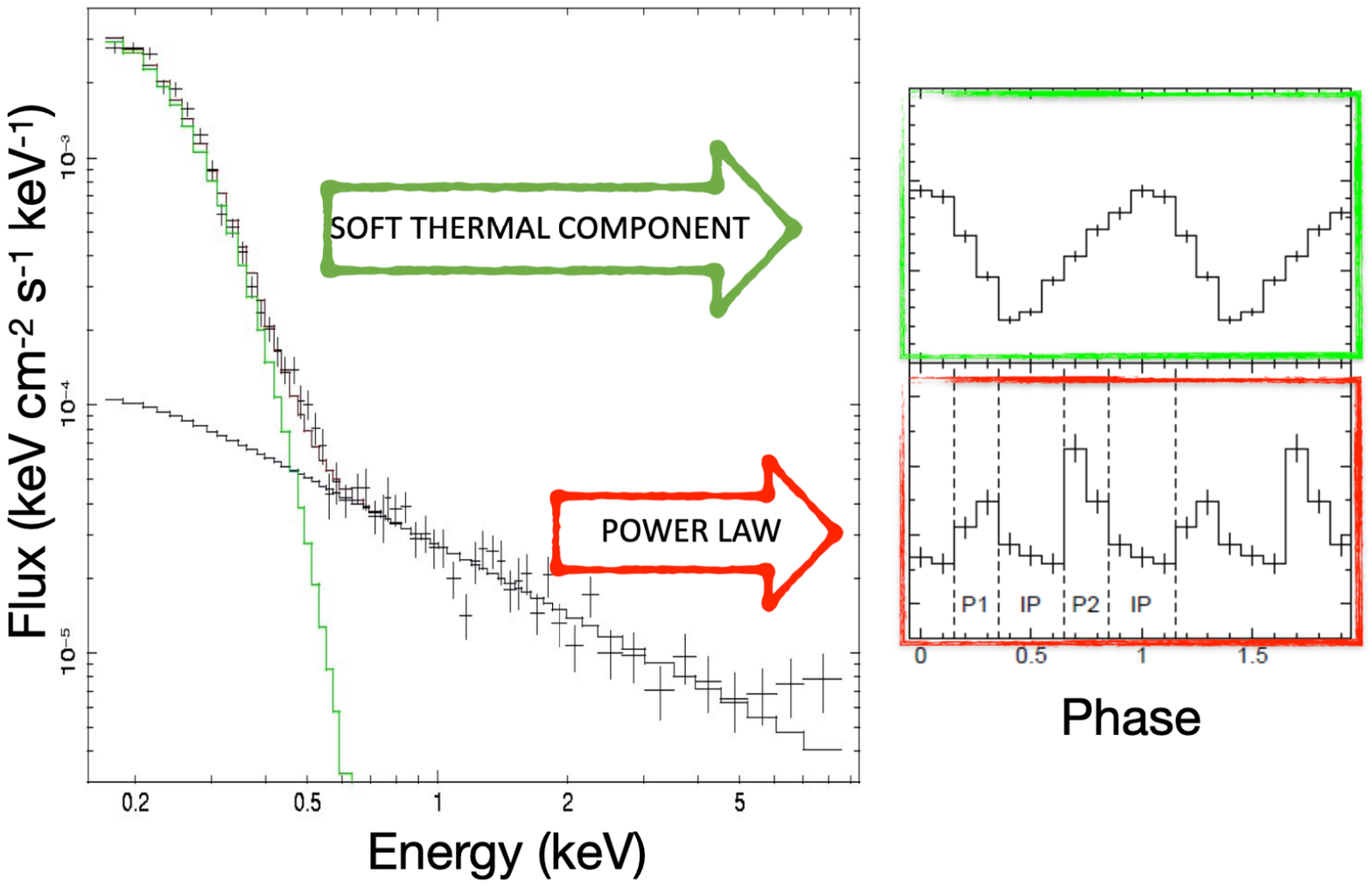}
\caption{X-ray spectrum (left panel) and pulse profiles in two energy ranges (right panel) of \hr . }
  \label{fig-spe}
\end{center}
\end{figure}

\begin{figure}[b]
\begin{center}
\includegraphics[angle=-90,width=11cm]{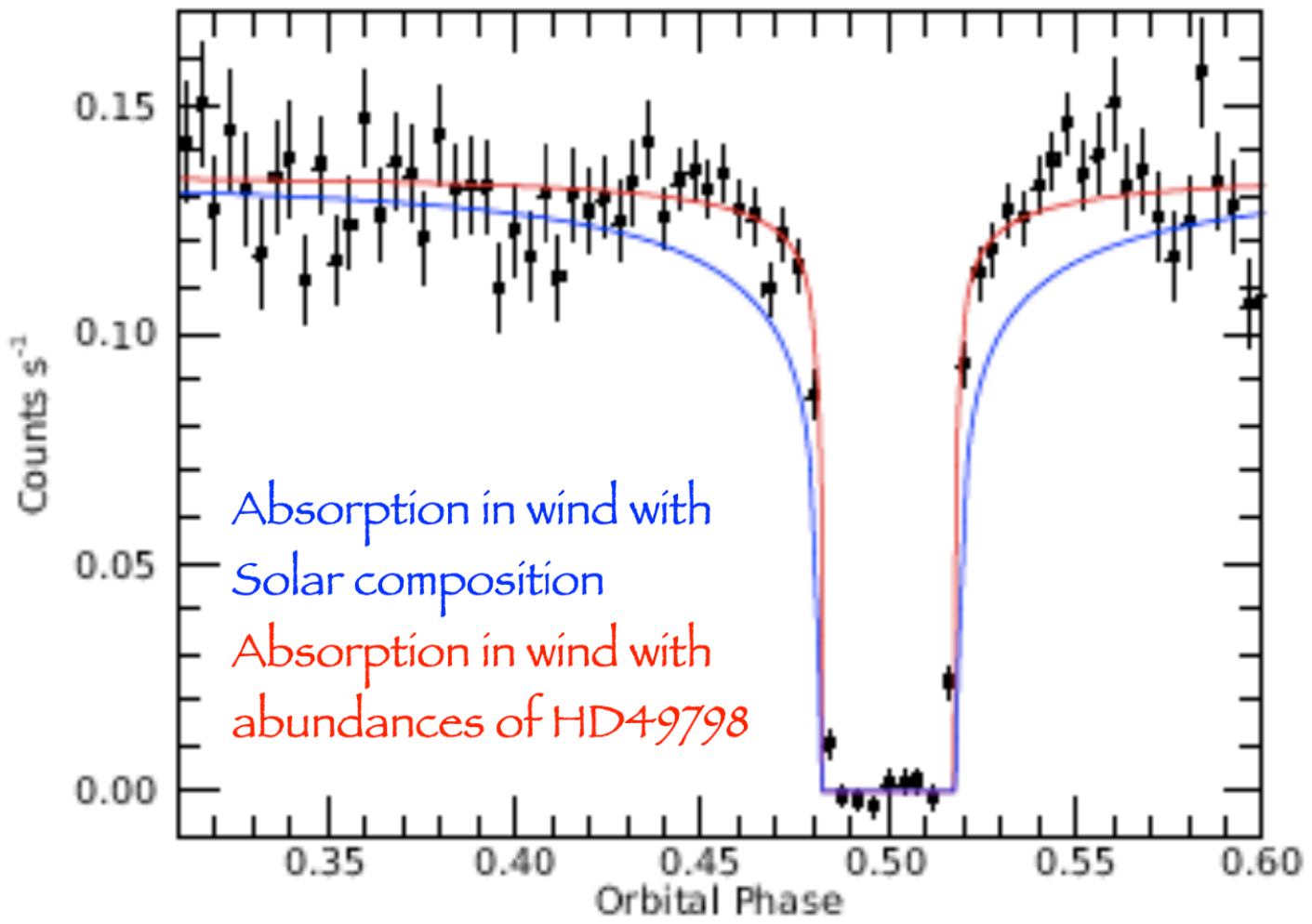}
\caption{Light curve of the X-ray eclipse compared with two models of absorption in the wind of \hd  .  The models have been computed by
integrating the  column density along the line of sight at different orbital phases, assuming solar abundances (blue line) and the abundances of \hd\ derived in Ref.~\citenum{krt19}.}
  \label{fig-ecl}
\end{center}
\end{figure}

The time derivative of the spin period  is negative\citep{mer16}, i.e. the WD is spinning-up at the very small, but remarkably steady rate of 72 ns per year\cite{rig23}. 
In fact, such a small spin-up rate could   be measured only by phase-connecting over almost 30 years all the available X-ray  observations. The implications of the source spin-up rate are discussed in Sec.~\ref{WDNS}. A search for the 13.2 s periodicity in the optical band\cite{mer11} gave negative results.
The X-ray pulse profile is energy-dependent, as shown in Fig.~\ref{fig-spe}. Below $\sim$0.5 keV it has a single broad peak, with a pulsed fraction of  $\sim$60\%, while above this energy it shows two peaks  with a smaller pulsed fraction.

Also from the spectral point of view it appears that the X-ray emission from \rx\ consists of two different components\cite{mer16,mer21}. The lower energy range is dominated by a soft, thermal-like emission, well described by a blackbody with temperature $kT$ = 31 eV (3.6$\times10^5$ K) and radius of emitting region $R_{BB}$=41 km,  while a power law with photon index $\Gamma$=1.9 and 0.3-10 keV flux of    $1.3\times10^{-13}$ erg cm$^{-2}$ s$^{-1}$  dominates above 0.5 keV (Fig~\ref{fig-spe}).  
The X-ray spectrum was also fitted\cite{mer21}  with a WD  atmosphere model  specifically computed for the  surface gravity of this massive WD and the composition of \hd\ (this is the expected composition of the   atmosphere, because the WD is accreting matter from the subdwarf's wind). This model  describes well the  soft spectral component, with a best fit with effective temperature $T_{eff}=2.2\times10^5$ K.  However,  the limited bandwidth and energy resolution of the available data did not allow us to obtain a univocal solution. In fact, only slightly worse fits were obtained with $T_{eff}=2.9\times10^5$ K and $T_{eff}=3.7\times10^5$ K. The most interesting result of this analysis is that the size of the emission region inferred with the WD atmosphere model ($\sim$1600 km, $\sim$400 km, and $\sim$66 km,  for the three values of $T_{eff}$) are much larger than that obtained using a simple blackbody to describe the thermal emission.

The soft X-ray emission is modulated near the eclipse ingress and egress by the absorption caused by the wind material (Fig.~\ref{fig-ecl}). These variations can be well modelled by adopting the correct composition of the wind \cite{krt19}.
Contrary to the typical behaviour of accretion-powered X-ray sources, the X-ray  luminosity has remained at the same level in all the observations,  except for a small flux variation (less than a factor  two) in the power law component,  in data taken 6 years apart \cite{mer21}.

 \section{White dwarf or neutron star?}
 \label{WDNS}

Both a NS and a WD are compatible with the measured mass and spin period\footnote{a period of 13.2 s, although the fastest observed in a WD, is still a factor $\sim$3 larger than the break-up value for such a massive WD\cite{cha87}.} of the compact companion of \hd .
The main arguments to discriminate between the two possibilities are based on three properties: 
\textit{i)} X-ray luminosity,
\textit{ii)} spin period evolution
\textit{iii)} size of X-ray emitting region.

 While initial estimates of the X-ray luminosity were affected by the uncertainties on the distance and on the spectral parameters,   the accurate Gaia parallax and the high quality  \xmm\ spectrum\citep{mer16,rig23}, imply a precise value of $L_X = (1.3\pm0.3)\times10^{32}$ erg s$^{-1}$.
 Given that the radius of \hd\  is much smaller than that of its Roche-lobe ($R_{RL}\sim$3 $\rsun$), accretion onto the compact object can occur only through capture of the stellar wind. In this case we can estimate an accretion rate of 
 %
%
$\dot{M} = \left(R\pdx{A}/2a\right)^2 \dot{M}\pdx{w}$, 
%
where $a$= 7.8 $\rsun$ is the orbital separation and $R\pdx{A}$ is the  Bondi-Hoyle accretion radius. The latter depends on the relative velocity between the wind and the accreting object, $v=(v_w^2+v_{orb}^2)^{1/2}\sim v_w$.  If this matter reaches the surface of a star with mass $M$ and radius $R$, we expect a luminosity

\begin{equation}
L\pdx{X}=\frac{GM}{R}\,\frac{R\pdx{A}^2}{4a^2} \dot{M}\pdx{w} = 
\frac{(GM)^3}{R~a^2~v_w^4} \dot{M}\pdx{w} 
 \\
 \label{eq:LNS}
 \end{equation}

\noindent
Taking  $\dot M\pdx{w}=2\times10^{-9}$  $\msun$  yr$^{-1}$,  $v_w$ = 1500 km s$^{-1}$, and  radii of 3000 km and 12 km, respectively  in the  WD and NS case, Eq.~\ref{eq:LNS}  gives
$L\pdx{X}\apx{WD}=1.2\times10^{31}$ erg s$^{-1}$ 
and
$L\pdx{X}\apx{NS}=3.0\times10^{33}$ erg s$^{-1}$, 
These luminosities are,  respectively,  lower and higher than the observed value.

\begin{figure}[b]
 \vspace*{-2.0 cm}
\begin{center}
\includegraphics[angle=-90,width=13cm]{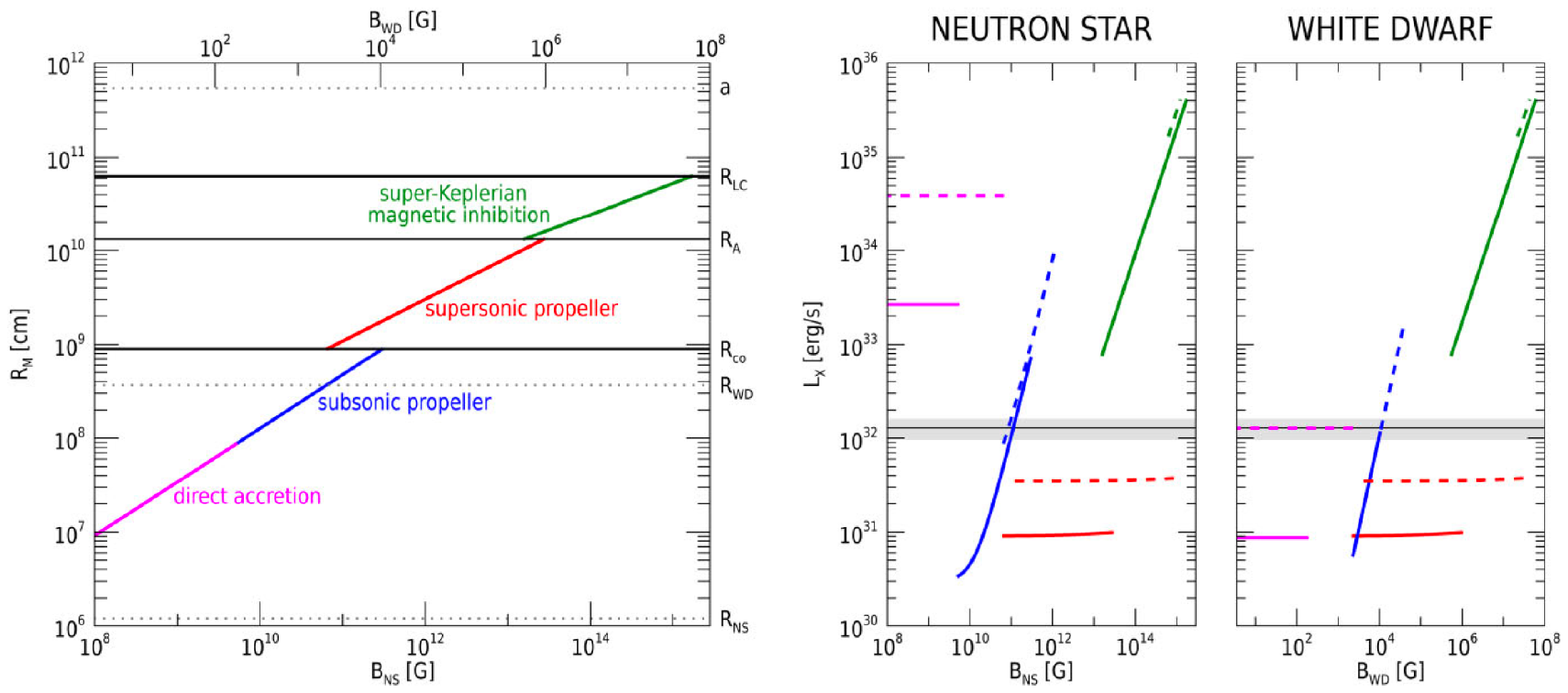}
 \vspace*{-1.0 cm}
\caption{-- (From Ref.~\citenum{rig23}) --
The left panel shows the magnetospheric radius as a function of magnetic field for a NS or a WD (bottom and top scale, respectively). The  different regimes depend on the relative dimensions of various characteristic radii:     the light-cylinder radius $R\pdx{LC}=cP/2\pi \simeq 6.3 \times10^{10}$\,cm, the accretion radius $R\pdx{A} = 2GM\pdx{X}/v_w^2 \simeq 1.3\times10^{10}$\,cm (for $v_w=1570$\,km\,s$^{-1}$), and the corotation radius $R\pdx{co}= (GM\pdx{X}P^2/4\pi^2)^{1/3} \simeq 9.0 \times10^{8}$\,cm. The orbital separation $a\simeq5.4\times 10^{11}$\,cm, 
    the WD radius $R\pdx{WD}=3580$\,km and the NS radius $R\pdx{NS}=12$\,km are also indicated.
  The two right panels show the  X-ray luminosities expected in the different regimes.  The observed value $L\pdx{X}= (1.3 \pm 0.3)\times 10^{32}\,$erg s$^{-1}$ is indicated (with its uncertainty as a grey horizontal band). Solid lines refer to $v_w=1570$\,km\,s$^{-1}$, while the dashed lines to $v_w=800$\,km\,s$^{-1}$.}
  \label{fig-reg}
\end{center}
\end{figure}

However, the presence of a rotating magnetosphere   that interacts with the accreting plasma can lead to more complex scenarios yielding  different luminosities\cite{lip92}. 
As discussed  in detail in Ref.~\citenum{rig23} and illustrated in Fig.~\ref{fig-reg}, a reduced NS luminosity, consistent with the observed value, can be attained only in the  sub-sonic propeller   regime  and if the NS has a magnetic field of $\sim$10$^{11}$ G. However, in this regime a  pulsed fraction much smaller than the observed one is expected and, more importantly,  it is impossible to explain the     spin-up. 
On the other hand, in case of a WD,  Eq.~\ref{eq:LNS} gives a luminosity in agreement with the observed one  for a wind velocity $v_w\sim800$ km s$^{-1}$. Although this values is smaller than the terminal wind velocity inferred from optical/UV studies (see Table~\ref{tab-par}),   it is possible that the wind velocity close to the WD is reduced due to the photoionisation produced by the X-ray emission\cite{krt18}.

A WD interpretation is also favoured by the time evolution of the spin period. Compact objects in X-ray binaries experience spin-up due to the angular momentum carried by the accreting material. However, both in the NS and WD case, the accretion rate in this system is too small to produce the observed spin-up rate, even in the most favourable condition of   accretion through a disk\footnote{The conditions for the formation of a disk are not satisfied in this system, therefore the   angular momentum accreted by the compact object is lower than the maximal value.}. Also the great stability of the spin-up rate over three decades, is at variance with the irregular variations typically seen in the period derivatives of wind-accreting systems.  A solution of this puzzle is naturally provided if the compact object is a relatively young WD, still in the radial contracting phase\cite{pop18}. The change in the moment of inertia for a WD of 1.2 $\msun$ and age between one and five millions of years gives a  spin period time derivative in full agreement with the observed one. Such a range of WD ages is consistent with the evolutionary scenarios proposed to explain  this binary system (see Sec.~\ref{evol}).

To discriminate between a WD and a NS, one can also consider the size of the emission region derived from the spectral fit of the soft spectral component, although this is clearly a model-dependent inference.
The blackbody model gives a radius of $\sim$40 km, while fits with a WD atmosphere model yield much larger radii    (Sec.~\ref{xray}). 
These dimensions are too large for a hot spot on the surface of a NS, but are reasonable values in the WD case.

 \section{Evolution}
 \label{evol}

The low H  abundance  indicates  that \hd\ is the stripped core of an initially much more massive and larger star.
Also the  N overabundance and the  low abundances of C and O confirm that its surface layers once belonged to the outer part of the
H-burning core of a massive star. Slightly different evolutionary scenarios  have been proposed to explain this system, that clearly passed through a common envelope phase.

 \hd\  could be the descendent  of a fairly massive progenitor that began to fill its Roche-lobe before He ignition and is currently in a core He-burning phase\cite{kud78,ibe93}.  
The mass of the He star can then be related\cite{ibe85}   to that of its progenitor by M$_{He}$=0.043 M$^{1.67}$, implying an initial mass of $\sim$8 $\msun$.
Alternatively,   \hd\ could consist of a degenerate CO core, surrounded by a helium envelope with a He-burning shell at its base\cite{bis97}. In this scenario, it would be the core of a progenitor of $\sim$5 $\msun$ that suffered  mass loss during a common envelope event when it was on the early asymptotic giant branch. Both scenarios imply that the WD in this system is much younger than those found in cataclysmic variables.

Also the  future evolution of this system, in which a massive WD is accreting He-rich material, is interesting and it has been discussed by several authors.
The current mass  transfer onto the WD is not changing significantly its mass, but it will greatly increase  when, about 65,000 yrs from now\cite{bro17},  \hd\ expands and fills  its Roche-lobe. This will produce a high X-ray luminosity,  possibly with the presence of bright outbursts and the formation of a super-soft X-ray source, and also possibly leading to  important variations in the WD mass.

Evolutionary computations\citep{wan10} indicate that  the WD will reach the Chandrasekhar   limit  after only a few 10$^4$ years of Roche-lobe overflow,
 during which $\sim$(5--6)$\times$10$^{-6}$ $\msun$ yr$^{-1}$ of
He-rich matter are steadily converted to C and O, while the unburned matter is ejected by the system
trough an optically thick wind.
The final fate of the WD depends on    other critical factors  such as,
e.g.,  its composition and final rotational velocity.

 If \hd\ hosts a CO WD, it could be the progenitor of an over-luminous type Ia supernova, since the fast rotation can
increase the mass stability limit above the value for non-rotating stars. Massive WDs are expected to have
an ONe composition, but again the high spin might play a role here, since it can lead to the formation of  CO WDs
even for high masses\cite{dom96}.
The fact that this system originated from a pair of relatively massive stars ($\sim$8--9 $\msun$) might suggest that it could
be the  progenitor of a type Ia supernova with a short delay time. However, the delay time might be considerably
longer if the explosion has to await that the WD spins down \cite{dis11}.

Considering its high mass, it is more likely that  the companion of \hd\ is an ONe WD. In this case an accretion-induced collapse might occur,
leading to the formation of a NS. The high spin rate and low magnetic field make
this WD an ideal progenitor of a millisecond pulsar.
This evolutionary scenario could provide an alternative path  for the direct formation of  millisecond pulsars, not involving the recycling of old pulsars in accreting low-mass X-ray binaries. 

The future evolution of this system in the case that the compact object is a NS has been explored in Ref.~\citenum{bro17}. According to these authors, \hr\  could pass in a phase of high accretion rate  appearing as an ultra luminous X-ray source\cite{kaa17}, and then produce either  a binary or an isolated millisecond pulsar.

 \begin{table}
\tbl{The fastest spinning white dwarfs}
{\begin{tabular}{@{}lcccc@{}}
\toprule
Name                                               & $P$ &  $\pdot$        &       WD mass   & Pulsation   \\
                                                   & [s] &  [s s$^{-1}$]       &       [$\msun$]  &    \\
 \colrule
        \hd\                                           & 13.2          & --2.3$\times10^{-15}$     &  1.22  & X-ray  \\ 
 LAMOST J024048.51$+$195226.9  \cite{pel22}    & 24.9          &      -- &  $>$0.7     &    opt.       \\ 
  CTCV J2056$-$3014    \cite{lop20,sal24}                     &   29.6            &   $<|2\times10^{-12}|$ & 0.7--1   & X-ray/opt.  \\
   AE Aqr     \cite{pat79}                                   &    33                 &  6$\times10^{-14}$   & 0.63 &   \\
     W1460 Her         \cite{ash20,pel21}        & 39                     &  $<|3\times10^{-14}|$  &  0.87  & UV/opt. \\
 \hline
WZ Sge  \cite{pat98,nuc14}                                      & 27.87  / 28.96  ? &  8$\times10^{-12}$  & 0.85 ? &    X-ray/opt.  \\
SDSS J093249.57$+$472523.0  \cite{twe24}                                      &  29.8 ? &   --  & 0.75 ?  &  UV  \\
\botrule
\end{tabular}}
\begin{tabnote}
The periodicities observed in the last two entries of the table are not yet securely associated to the WD rotation and
the masses of these WDs have not been determined. \\
\end{tabnote}
\label{tab-WD}
\end{table}
 
 \section{Fast rotation and magnetic field}

The  companion of \hd\ is the WD with the shortest spin period currently known. 
A few other fast rotating WDs have been discovered in recent years (see Table ~\ref{tab-WD}).  
All of them are in cataclysmic variables,  old systems in which the WD has been spun-up during a long phase of disk accretion.
On the other hand,  the origin of the rapid rotation of the  \hd\ companion is unclear.
\hr\ is a young system, originating from intermediate mass stars,  that  passed through a  common-envelope phase\cite{ibe85}.
The current  transfer of angular momentum is very small.  
Unless the  WD was directly born with a short period, the spin-up must have occurred either before or during the common envelope phase.
Given the  short duration and complex dynamics of this phase, it 
is more likely that  the spin-up occurred  when the subdwarf was close to fill its Roche-lobe,
just before the ensuing of the common-envelope phase and with the formation of an accretion disk.

If the accretion disk extends down to the star's surface, as it is expected for a weakly magnetised WD, short  periods  can be easily  
reached thanks to the high specific angular momentum carried by the accreting matter.  
Interestingly, some indication on the strength of the magnetic field can be derived considering the constraints illustrated in Fig.~\ref{fig-reg} for the different
regimes of interaction between the rotating magnetosphere and the surrounding plasma. 
As discussed in Refs.~\citenum{mer09,rig23}, the strong X-ray pulsations and the observed luminosity imply a magnetic dipole field below  a few 10$^3$ G,  otherwise the WD  would be in the propeller state.  
If the WD magnetic field was so weak also during the  phase of rapid spin-up due to (near) Roche-lobe accretion, it should have been spun up to a period even shorter than the current one.

  \section{Conclusions}
 \label{conc}
 
Many binaries composed of a WD and a hot subdwarf  of B spectral type are known\cite{gei11,gei23}, but no accretion powered X-rays  have been detected in these systems\cite{mer11b,mer14,mer22,des22}. Thus, \hr\ is the only known accretion-powered X-ray binary in which the mass donor is a hot subdwarf.  Furthermore, both components of this binary do not share the typical properties of their respective classes:
the WD is particularly massive and has a high spin frequency (though it is weakly magnetic),   the hot subdwarf has a high luminosity, a relatively low surface gravity and
peculiar abundances.  These properties make \hr\ particularly interesting, not only in the context of stellar evolution models,  but  also  for the study of hot subdwarfs and WDs in general. For example, it is possibly the first observed example of a young WD  still  in the   phase of contraction. It is also a laboratory to investigate  the  weak stellar winds of hot subdwarfs, since the orbiting WD acts as a probe of the wind properties. 
 Future multi-wavelength observations can bring very useful information and help to solve some of the still open questions, such as, e.g., the origin of the fast spin and the peculiar spectral/timing properties of the X-ray emission.  In a more distant future, the presence of a fast rotating massive WD might also lead to the detection of gravitational waves caused by     asymmetries in the mass distribution caused by accretion or by magnetic deformation\cite{sou24}.

\section*{Acknowledgments}
I thank all the colleagues and friends with whom I collaborated over many years in the  study of this interesting binary.  I acknowledge financial support from INAF through the Large Program Grant  ``Magnetars''.


 \bibliographystyle{ws-procs961x669}

 \vfill
\pagebreak
 
\end{document}